# From Compliance to Impact: Tracing the Transformation of an Organizational Security Awareness Program

Julie M. Haney
National Institute of Standards and Technology

Wayne Lutters
University of Maryland College Park

**Abstract**
There is a growing recognition of the need for a transformation from organizational security awareness programs focused on compliance – measured by training completion rates – to those resulting in behavior change. However, few prior studies have begun to unpack the organizational practices of the security awareness teams tasked with executing program transformation. We conducted a year-long case study of a security awareness program in a United States (U.S.) government agency, collecting data via field observations, interviews, and documents. Our findings reveal the challenges and practices involved in the progression of a security awareness program from being compliance-focused to emphasizing impact on workforce attitudes and behaviors. We uniquely capture transformational organizational security awareness practices in action via a longitudinal study involving multiple workforce perspectives. Our study insights can serve as a resource for other security awareness programs and workforce development initiatives aimed at better defining the security awareness work role.

## 1 Introduction

Compliance – aligning organizational processes and programs with external rules and standards – is a significant driver for security (cybersecurity) programs in various sectors, such as healthcare, government, and financial services. Compliance can be helpful in setting a minimum bar of security expectations for an organization. Training employees for policy compliance influences their awareness but has minimal impact on promoting actual compliant behavior (Stevens et al., 2020). Addressing this gap between knowledge of and practice of, is the purview of security awareness training.

Security awareness training plays an important role in helping organizational employees achieve a common understanding of security threats and acceptable security-related actions (Wilson and Hash, 2003). Various public and private industry sectors recognize the importance of this role by requiring or recommending annual security awareness training. For example, U.S. government agencies implement training mandated in the Federal Security Modernization Act (FISMA) (2014). The European Union's General Data Protection Regulation (2016) encourages organizations to provide similar training.

These training requirements intend that compliance will result in positive impacts on workforce security behaviors, thus improving the overall security posture of organizations. While training can be helpful, superficial compliance metrics indicating number of employees completing the training tell little of deeper impact: how security behaviors, understanding, and attitudes have positively changed. Indeed, prior literature and industry surveys revealed that security awareness programs often fall short in changing behaviors (Bada *et al.*, 2019). Employees continue to fall prey to cyber attacks, as exemplified by the Office of Management and Budget (2021) report stating that 53% of U.S. Government cyber incidents in 2020 resulted from employees violating acceptable usage policies or succumbing to email attacks.

Considering these sobering observations, there is a growing recognition of the need for transformation from security awareness programs that are merely compliance-focused to those facilitating behavior change. While studies have examined the impact of security awareness efforts from the *end user* perspective (Khando et al. 2021; Tschakert and Ngamsuriyaroj, 2019), few have begun to unpack the *organizational practices* of security awareness professionals tasked with executing program transformation. To address this shortfall, we conducted longitudinal, case study research of a security awareness team and program at a U.S. Government agency. Through interviews, field observations, and document analysis, we found a deliberate progression of the program from being focused on training compliance to becoming dedicated to achieving employee empowerment and behavior change.

Our study makes several contributions. We uniquely capture the transformation of security awareness practices within an organizational context from the perspectives of multiple stakeholders (security awareness team members, their managers, and employees), instead of the end-user perspective common in other studies. In doing so, we confirm and extend the research body of knowledge into a real-world context. Furthermore, our findings have applicability for other security awareness programs by allowing readers to witness how one organization re-examined and evolved their program beyond a compliance focus. While specific tactics may not be appropriate in all organizations, the considerations and strategies employed by our study's awareness team can serve to inform how other practitioners might approach maturing their own programs. We also contribute to efforts to define a dedicated security awareness work role by confirming the skills needed by these professionals.

## 2 Related Work

Our case study is focused on understanding how security awareness professionals develop, execute, and transform their security awareness programs. Because our study explores the perspective of the awareness team, we do not fully address literature on security awareness impacts and perceptions from the end user perspective.

### 2.1 Organizational Security Awareness Programs

#### 2.1.1 Security Awareness Approaches and Challenges

Hänsch and Benenson (2014) posited three dimensions of security awareness commonly identified in research literature. *Perception* is gaining knowledge of the existence of a security threats. *Protection* involves users knowing what measures they must take to counter these threats. *Behavior* refers to users actively and effectively taking steps to reduce security incidents.

The cornerstone of many security awareness programs is annual training, most often conducted online. Beyond this training, programs may implement other awareness activities and communications throughout the year, for example, speaker events, newsletters, emails, and even novel approaches such as virtual reality and escape rooms (Haney et al., 2022). These additional activities can reinforce learning while adapting to new security threats, policies, and processes as they arise.

Although viewed as important for organizational security, security awareness programs may face several difficulties. Training often has a poor reputation of being boring and ineffectual (Reeves and Delfabbro, 2021). Additionally, programs may fail to provide engaging materials that motivate employees to practice good security habits, have unrealistic expectations of what employees can do, and rely on punitive, fear-based tactics that may not have a positive impact on sustainable behavior change (Pedley, *et al.*, 2020; Bada *et al.*, 2019; Alshaikh, 2018; Renaud and Dupuis, 2019).

To counter these challenges, researchers made recommendations for security awareness approaches. Based on surveys and a literature review of security behavioral theories, Moody *et al.* (2018)

developed the Unified Model of Information Security Policy Compliance (UMISPC) to guide security awareness efforts. The model illustrates the impacts of *habit*, *fear* (influenced by threats and response self-efficacy), and the *role values* (tasks and nature of work roles) on employee intentions to follow information security policies. Others took a practical approach, recommending specific awareness strategies. Bada *et al.* (2019) encouraged programs to disseminate simple, consistent guidance that promotes a sense of self-efficacy and continuously provide feedback and training refreshers to help make security a habit. Alshaikh *et al.* (2018) recommended engaging employees by: communicating how security impacts the organization; building trust and relationships with the workforce; soliciting employee feedback; relating security to employees' personal lives; and implementing creative types of training. Others (Bauer *et al.*, 2017; Abawajy, 2014; Korpela, 2015), suggested that organizations find varied ways to communicate security awareness and tailor communications since they cannot expect a single approach to resonate across the entire workforce. For example, serious games, in which participants learn about security concepts then put them directly into practice, have been found to be effective in raising awareness and teaching security techniques (Hart *et al.*, 2020; Khando *et al.*, 2021; Shostack, 2023).

The field work presented in this paper builds on this previous research to explore the challenges of one particular security awareness program and how the awareness team overcame these difficulties in practice.

*2.1.2 Measures of Effectiveness*

Measuring program success is a critical, but challenging, aspect of security awareness programs, with few programs making a concerted effort. The lack of measurement may in part be due to organizations' emphasis on compliance to training policies (e.g., FISMA), represented by training completion rates, when the goal should be behavior change (Fertig *et al.*, 2020; Bada *et al.*, 2019). For a holistic assessment, organizations can instead use a combination of measures of effectiveness, including: number and kinds of security incidents related to training topics; user-initiated incident reporting; phishing simulation click rates; engagement with security awareness materials (e.g., represented by number of views); and feedback from stakeholders via surveys and interviews (Alshaikh *et al.*, 2018; Fertig *et al.*, 2020). Chaudhury *et al.* (2022) developed a security awareness metrics framework that included the following indicators: *impact* (changes in security knowledge, attitude, and behavior); *sustainability* (value-added and impact on organizational policies and processes); *accessibility* (relevance, quality, reachability, and usability of awareness materials); and *monitoring* (workforce and leadership interest and participation). Furthermore, the security training institute SANS (2021) found that organizations that assess their own programs against peers tend to have greater leadership support for security awareness training, and, therefore, more success. One such benchmark is the five-level Security Awareness Maturity Model (SANS, 2018).

Beyond self-reports and researcher recommendations, our case study demonstrates the use and evolution of various measures of effectiveness as observed in practice.

*2.1.3 Security Awareness Professionals*

Security awareness professionals focus on raising awareness of and advocating security practices to the general workforce within an organization. Surveys of these professionals revealed that most perform security awareness duties on a part-time basis without a job title that reflects their duties, and they may lack sufficient professional skills (e.g., communications) that were viewed as essential in security awareness roles (Haney et al., 2022; SANS, 2021; Woelk, 2015). Stewart and Lacey (2011) described a significant issue in that the technical specialists who are often given security awareness responsibilities mistakenly believe that providing a "broadcast of facts" will result in behavior change and fail to consider how communications should be tailored to their audiences. Bada *et al.* (2019) highlighted competency issues when they suggested that the lack of understanding that awareness is a unique discipline leads to ill-prepared awareness professionals.

While prior studies relied solely on self-report data, our study used multiple data collection methodologies to observe security professional practices in action.

### 2.2  *Organizational Security Influencers*

As a foundation of our research, we look to prior work related to workers who, like security awareness professionals, attempt to influence others' behaviors within organizations: risk communicators, change agents, and cybersecurity advocates. Security awareness professionals are risk communicators since they need to effectively communicate security risk, with a goal of enabling positive security behavior. The risk communication research body of knowledge, e.g., (Covello, 1997; Nurse et al., 2011; Slovic, 1987), identifies desirable practices of communicators, including: keeping communications simple, but specific and engaging; customizing information to target audiences; and assisting people in seeing the consequences of their decisions.

Those in security awareness roles may also be considered a type of change agent, described in Diffusion of Innovations (DOI) Theory as someone who actively influences clients' adoption decisions (Rogers, 2003; Markus and Benjamin, 1996). In facilitating adoption, change agents focus on the betterment of the organization, team with others, employ strong communication skills, and practice context-dependent adaptation.

Haney and Lutters (2018; 2021) defined cybersecurity advocates as security professionals who actively promote and facilitate the adoption of security best practices and technologies as an integral component of their jobs. Advocates are deeply passionate about the work as they see security having the potential for significant individual and societal impacts. In advocacy jobs, technical literacy, while necessary, is often regarded as less important than professional skills – interpersonal skills, communication, context awareness, creativity, and flexibility. Therefore, multi-disciplinary advocacy teams are viewed as beneficial in developing and disseminating engaging security guidance. The study also uncovered advocates' work practices, revealing that, similar to risk communicators and change agents, they strive to empower their audience by providing practical recommendations in an understandable language, employing engaging communication techniques, and incentivizing positive behaviors.

In this paper, we apply and extend these prior works into the security awareness domain by studying risk communication, change agentry, and security advocacy *in practice*. Specifically, while the Haney and Lutters study began to form a picture of the work of cybersecurity advocates by identifying what advocates *say* they do, our case study extends this work by providing a better understanding of what one type of advocate (security awareness professionals) *actually do* in a real-world setting.

## 3  Methodology

We conducted a single case study of the security awareness team and program in a U.S. Government agency over the course of one year. The study protocol was approved by our Institutional Review Board. The agency's Chief Information Officer provided a signed letter of support for study participation.

### 3.1  *Selection of Case Study Method*
While positivist research aims to predict, control, and generalize through the collection of quantitative data, our research goals to describe, understand, and interpret the processes and transformation of a security awareness program aligned with a constructivist approach (Merriam and Tisdell, 2016). Therefore, we made the careful decision to conduct a *qualitative* case study.

A case study is a bounded, in-depth investigation of an object of study ("the case") to capture its complexity. The case study methodology has been a staple of social science research for decades and is particularly appropriate when in-depth explanation of a phenomenon is required (Zainal, 2007). It typically involves the use of multiple data collection and analysis methods, allowing for triangulation to strengthen interpretations (Baxter and Jack, 2008; Yin, 1999). Eisenhart (1989) identified three possible aims of case studies: providing description, testing theory, and generating theory. The case study is particularly suited to our research goals since it attempts to both confirm prior findings (testing theory) and yield insights into security awareness through a holistic, detailed understanding of a program in a real-world context (providing description).

Practices surrounding security are some of the most closely guarded organizational behaviors, for understandable reason. Opportunities to gain deep participant observer access to these programs are rare, let alone having longitudinal access to organizations that are not your own, as evidenced by a lack of such studies in current literature. A strength of this study is precisely this kind of access – affording a robust view of security awareness work in action.

The object of the case study was the security awareness team and program at a U.S. Government agency (referred to as Agency Q). We selected a case that was theoretically useful and allowed us to replicate or extend prior theories and findings on security awareness and advocacy. In addition, due to a preponderance of government security mandates, government agencies tend to be compliance-oriented, offering an opportunity to explore how an agency balances security awareness compliance pressure with achieving actual impact. Additionally, the program was in an active state of transformation we could observe and describe real-time.

### *3.2 Data Collection*

To examine security awareness from different perspectives, we took a multi-methods data collection approach involving interviews, field observations, and reviews of both qualitative and quantitative documents to triangulate evidence. We collected data via 14 on-site visits and multiple email exchanges and phone conversations spread out over the course of one year. Our collection methods were responsive to the context (Eisenhart 1989), adjusting to accommodate new awareness events as they arose and new data sources as we became aware over the course of our investigation.

### *3.2.1 Interviews*

We interviewed the entire security awareness team: the program lead (a career civil servant) and two contracted staff. Prior to being interviewed, they completed a short demographic survey capturing education and work experience. Since our study extends prior work on cybersecurity advocacy into a specific, observable context (organizational security awareness efforts), our interview questions were initially based on the protocol used in the Haney and Lutters (2018) interview study with modifications to capture organizational details and account for the security awareness focus. Questions covered work practices, approaches, motivations, and challenges. The team lead was asked additional questions to obtain details about team composition and security awareness efforts. The lead's interview lasted over an hour, and the two contractor interviews were 30 minutes each.

To obtain a management perspective, we interviewed two managers in the team's chain-of-command: the Chief Information Security Officer (CISO) (interview time 20 minutes), and the direct supervisor of the team lead (45 minutes). We asked them about their views of the security awareness program, how program resource decisions were made, thoughts about security challenges to the organization, their roles, and experience at Agency Q (both had about 10 years).

We also interviewed nine agency employees who were "consumers" of the awareness program's events and materials. The awareness team selected and recruited the employees via email and face-to-face contact based on our request to include individuals in both information technology (IT) and non-IT roles. Employee interviews lasted about 30 minutes. Employees were asked about their views of the security awareness program, the security information they received, and the impact of that information on their behaviors. We also asked how long they had been at Agency Q, with responses ranging from 11 – 30+ years. Five employees were in non-IT roles. Employees also assessed their own security familiarity, with three indicating low familiarity, three moderate, and three high. Table 1 provides employee demographics.

Interviews were audio-recorded and transcribed. Data were stored without personal identifiers, rather with a participant code. Participant codes for security awareness team members are indicated with an "S" (S1-3), managers with an "M" (M1-2), and employees with an "E" (E1-9).

| ID | Role | # Years at Agency | IT Familiarity | Security Familiarity |
|---|---|---|---|---|
| E1 | Research librarian | 12 | Moderate | Moderate |
| E2 | Manager of an information security organization | 12 | High | High |
| E3 | Team leader of a research and technology resources organization | 14 | Moderate | Moderate |
| E4 | Engineer | 12 | Low | Low |
| E5 | Attorney | 30+ | Low | Low |
| E6 | Attorney | 12 | Low | Low |
| E7 | IT specialist, information systems security officer | 11 | Moderate | High |
| E8 | Information systems security officer | 17 | Moderate | High |
| E9 | IT specialist | 13 | Moderate | Moderate |

*Table 1.* Demographics of interviewed employees

### 3.2.2 Field Observations of Security Awareness Events

Field observations are a commonly-used, valuable research tool in gathering first-hand data, providing specific events to be used as reference points, and helping to triangulate emerging findings (Merriam and Tisdell, 2016). The security awareness team organized three types of events throughout the year: lunchtime events, security days, and security officer forums (described later). To observe the team's approaches in action, we captured a complete annual cycle of events as a participant-observer, attending at least one of each common event type: three lunchtime events, one security officer forum, and two security days. Detailed field notes taken during the events documented techniques, the security information presented, and audience engagement.

### 3.2.3 Additional Data Collection

We collected over 25 electronic and physical copies of security awareness materials disseminated during security awareness events and campaigns. The team also provided two other types of documents for analysis: 1) 20 post-event reports from the previous two years and year of the study and 2) six feedback survey reports from events during the same timeframe. These reports provided further confirmation of observational findings.

As emerging themes and additional questions arose, other data were collected as needed via email exchanges or phone conversations with the team. For example, one phone conversation revolved around learning more about simulated phishing exercises.

### 3.3 Data Analysis

Analysis of multiple data sources allowed us to find overlapping themes and identify the "why" behind identified relationships. Data analysis of interviews and event field notes was initially guided by a subset of deductive codes based on the cybersecurity advocates study (Haney and Lutters, 2018), with inductive coding using Grounded Theory-informed methods (Corbin and Strauss, 2015) employed as new data labels emerged. Both authors independently coded a sample of interviews: all team interviews, one manager interview, and two employee interviews. We then met to discuss the applicability of the deductive codes, reorganize codes based on differences in the case study as compared to the previous interview study, and suggest new codes as appropriate to construct a final codebook. The first author than used the codebook to deductively code the remaining interviews and field notes. We regularly met to discuss emerging themes and our interpretations of the data.

We also reviewed the other provided documents to glean which security topics the team deemed important to distribute and how they leveraged third party resources. A review of post-event reports provided a deeper understanding of past security awareness events, including attendance, topics, and how events were portrayed to agency leadership. Event feedback surveys offered insight into how the events were viewed by attendees and how feedback contributed to subsequent events.

### *3.4  Researcher Positionality and Limitations*

In qualitative research, the researcher serves as the primary instrument for data collection and analysis (Merriam and Tisdell, 2016). Therefore, the positionality of the authors is important to note. The first author has a background in cybersecurity and human-centered computing as both a practitioner and a researcher working in the U.S. Government. The second author has an academic research background in information and computer science and specializes in field studies of IT-mediated work from a socio-technical perspective.

These positionalities and related experiences explicitly surfaced throughout data analysis discussions and potentially impacted the interpretation of data. They often served as an advantage, for example, facilitating the first author's trust-building with Agency Q staff and understanding of common security terminology used by the awareness team. Any undue bias introduced by these positionalities was mitigated by a rigorous data analysis process, as well as full-team data analysis discussions during which assumptions were identified and deliberated.

We also note several potential limitations of our study. The case study, which was valuable for exploring security awareness in practice, was bounded in scope as it focused on a team working in the government sector. Therefore, we cannot generalize findings to all employment sectors and types of awareness programs. However, given corroborating results in prior research and other projects focused on security awareness training and professionals, we believe the findings from our study may be transferable to similar contexts.

Additionally, the security awareness team's involvement in the recruitment of employees to interview may have resulted in a biased sample. Although we requested a diverse sample of work roles, employees more familiar with and favorable towards the awareness program and security may have agreed to be interviewed, with 4 of 9 employee participants in a security role. However, without our own access to internal email addresses, the sample selection was the most practical avenue given resource constraints that prevented the awareness team from assisting with large-scale recruitment. Although the employee interviews did provide interesting insights, this limited sample was not enough to make general inferences across the entire workforce, and we gained little understanding of those who chose not to participate in awareness events or those who attended and were unsatisfied.

## 4   Case Study Context

As context for our findings related to security awareness program evolution, this section provides thick descriptions of the security awareness team, initiatives, and challenges.

*4.1 Agency Q*

Since security practices are some of the most sensitive information that an organization holds, gaining extended insider access to this is exceptionally rare. The access provided to us was wide-reaching, but still constrained by a disclosure agreement to preserve anonymity. Here we share as much information about the organizational context as we are able within these constraints.

Agency Q is a medium-sized U.S. government agency on the scale of 5000 government and contract employees with a mission focused on producing guidance and regulations for critical infrastructure stakeholders. Most employees are stationed at agency headquarters, with smaller contingents in regional offices throughout the country. Employee roles are diverse, ranging for scientists and engineers to IT professionals to business and support personnel.

The agency deploys Microsoft Windows and Apple MacOS computers and Android and iOS smart phones for official use by employees. At the time of the study, employees were authorized to telework one day per week using government-owned resources. Sensitive information is routinely produced, disseminated, and stored by agency employees.

Agency Q fell prey to three different coordinated spear-phishing attacks several years prior to the study. The attacks resulted in about a dozen employees disclosing their account and password information and two others infecting their computers with malware. Afterwards, there was a concerted effort to educate the workforce on detecting phishing emails and appropriately using email.

*4.2 Security Awareness Team*

The security awareness team was in a security oversight organization under the Office of the Chief Information Officer (CIO). The program dated back at least 10 years prior, at which time it focused on annual training compliance with few in-person events. The current team assumed responsibility for the program three years prior to the start of our study and had expanded the program to three primary responsibilities: 1) implementing and tracking the completion of online, mandatory, annual security training; 2) planning and executing initiatives to increase employee awareness of security issues and appropriate actions, and 3) managing role-based training required for employees with a security role (not a focus of our study).

The security awareness lead (S1) had over 30 years of IT and security experience with degrees in business administration and computer programming. S1 spent 50% of work time leading the security awareness program, which was viewed as "*the most important thing I'm doing right now*." S1 also worked with information system owners on system authorizations and served as a subject matter expert for other government security initiatives.

The other team members (S2 and S3) were contractors supporting the program full-time. Neither had formal backgrounds in a technical field but brought business skills that enabled them to be heavily involved in the creative and logistical aspects of event planning. S1 appreciated the discipline diversity within their team, admitting that, although S1 had the security and technical knowledge, "*I am not a subject matter expert in the marketing and…the organizational skills. I need [S2 and S3] to do that. They're also brilliant when it comes to the ideas and how we're going to make things happen.*"

*4.3 Initiatives*

The team implemented several initiatives (activities) throughout the year: three types of synchronous, in-person events (lunchtime events, security officer forums, and security days) and two types of asynchronous activities (campaigns and phishing exercises). Table 2 provides an overview of these. Most initiatives had themes. For example, a "Safety Tech Check" event helped attendees get the most out of their smartphones, laptops, and tablets while operating them safely and protecting personal information.

The team developed their ideas during brainstorming sessions informed by their own explorations of security awareness techniques and topics. Their expertise developed from via attendance at external security events, information exchanges with other security awareness professionals, vendor security training, self-study, current events, and subscriptions to online security forums. The team also routinely considered agency mission priorities, workforce feedback, and organizational security incidents and risks (see section 5.1).

| Initiative | Description | Target Audience & Typical Attendance | Number per year |
|---|---|---|---|
| Lunchtime Events | These were informal, drop-in events held in an exhibit area located in a main thoroughfare. Typical events had an information fair atmosphere with multiple tables staffed by agency employees, representatives from community organizations, or other government agencies. However, alternate, performance-based event formats had recently emerged for these events. | General workforce<br><br>~100 employees | 4-5 |
| Security Officer Forums | The 2.5-hour forums focused on providing information to help security staff in their roles. Forums were held live but were broadcast to the those in remote locations and recorded for later viewing. Forums typically featured security leaders providing agency security updates and 4-5 talks by internal or external speakers. | Agency staff with security roles (but anyone can attend)<br><br>~100 employees | 2 |
| Security Days | The half-day events took place in a large auditorium and were aimed at all agency employees with a goal of providing security and technology information applicable to at both home and work. There were typically four talks during the events. The events were remotely broadcast with recordings and presentation slides posted for later viewing. | General workforce<br><br>~200-300 employees | 2 |
| Campaigns | Focused campaigns and security awareness material distribution addressed specific security issues emerging as agency challenges. Branded campaign posters were displayed throughout agency buildings, and handouts outlining mitigative actions were distributed to the workforce. | General workforce | As needed |
| Phishing Exercises | Simulated phishing attacks to test the workforce's susceptibility to phishing emails and raise awareness. The security awareness team set the strategy for these exercises, but a contractor executed the exercises and collected statistics on click rates. | General workforce | 4 |

*Table 2*. Agency Q's security awareness initiatives.

### 4.4 Challenges

Interviews revealed challenges that hindered security awareness efforts and the agency workforce's willingness and ability to practice good security habits. To set the context for exploring how the team tried to overcome these challenges, the most frequently mentioned issues are described here.

### 4.4.1 Lack of Buy-in

The team was challenged by employees not in IT or security roles who did not recognize the value of security, how it related to them, or their own security responsibilities. S1 discussed difficulties convincing non-technical staff to attend awareness events because *"we're still in our silos."* An engineer said that security was not a concern in their group because *"we don't do that for a living" (E4)*. Security may be de-prioritized as employees are *"busy doing their work" (M2)*. Several interview participants discussed the need for better alignment between the security awareness program and the agency's mission elements to facilitate understanding of the connection to security.

Cognitive biases may also affect buy-in. For example, employees may be overly optimistic in believing they would not be a victim of a security attack because they do not do anything important enough to be targeted. A security officer felt people may not *"take it seriously until it personally affects them" (E7)*.

Lack of support for security training was especially problematic when some agency leaders *"see it as a nuisance" (S1)*, rather than setting an example for the workforce. S1 commented, *"I just happened to have a training course with some leadership in the mission office. And they're like 'Why do I have to take this course? Why do I care about cybersecurity?'" (S1)*. With this attitude, managers may not support employees attending events because they feel "*I can't afford to send my people there…They need to be here at their desk" (S1)*.

### 4.4.2 Compliance-driven Training

Compliance to government security mandates, such as those specific to employee security awareness, is an important metric for government organizations, including Agency Q. However, the security awareness team recognized that *"just because you're compliant doesn't mean that it's an effective program" (S1)*. Compliance metrics failed to show how impactful the training was in changing workforce behavior: *"Everyone will check the box saying 'Yeah, we're 100% trained.'…Even if you are, what good is it if you don't have people applying what they learned?" (S1)*.

Annual security awareness training might be the same every year and was viewed by employees as burdensome. S1 commented, *"People do it because they have to. You have the online training where it's like 'Click, click, click, click, done'…rather than paying attention to the words they're seeing on the screen" (S1)*. The team observed that employees completed training to avoid punitive measures so "*their supervisor will get off their back" (S3)*, rather than as a learning opportunity.

### 4.4.3 Resources

In addition to organizational and staff challenges, the team was faced with resource shortages. The program budget was just enough to pay for the two supporting contractors. Without additional resources, the team was not able to expand their efforts, as confirmed by the CISO: *"I don't know how much more we can do to increase our activities with the budget pressure that we've got" (M1)*.

## 5 Findings

Over the prior three years, the security awareness team made a concerted effort to transform their program with a renewed focus on engaging and empowering employees to make sound security decisions rather than simply on enforcing compliance. In this section, we detail findings based on our observations of the strategies and activities that demonstrated the program's ongoing evolution.

## 5.1 Evolving to a Cybersecurity Advocate Role

Our exploration of program progression towards behavior change is, in part, illustrated by how the agency's awareness team members actively took on a cybersecurity advocate role (Haney and Lutters, 2018).

### 5.1.1 Building Trust

The cornerstone of cybersecurity advocacy is the ability to establish trust with the audience. To build technical credibility, the awareness team stayed updated on security topics through formal training, self-study, subscriptions to mailing lists, and being *"on alert with current [security] events or anything in the news" (S3)*.

Beyond domain knowledge, the team demonstrated interpersonal skills critical for building relationships and trust. S1 felt that, for security to resonate on an individual level, a personal touch is required, so preferred face-to-face interactions whenever possible. During security awareness events, we directly observed that the team members were enthusiastic and positive, projecting their service-oriented motivation to help people navigate the complexities of security *"so they can just be aware of what's going on and what they can do to make their life easier and protect themselves" (S1)*. For example, during a lunchtime event, we observed S3 facilitating a cybersecurity trivia game. They praised contestants for correct answers. If participants answered incorrectly, S3 was non-judgmental and explained the answers in an easy-to-understand way. Attendees visibly responded positively. Employees also expressed confidence in both the team and the information in the interviews, further illustrating the team's success in establishing trust. For example, an employee said that team members *"are knowledgeable…They're the experts…They are incredibly energetic" (E3)*. Another thought the team was approachable: *"They're friendly. They're open to suggestions. They are open to questions" (E9)*.

### 5.1.2 Effectively Communicating

In cybersecurity advocacy, communication skills are essential. To communicate effectively, the team had a solid understanding of organizational context, which aided them in selecting appropriate communication mechanisms and styles. For example, at the observed security officer forum, S1 injected organizational context when talking about cybersecurity role-based training, including information about the federal directives mandating the training, who it applies to within the organization, why it is important to the organization, and how agency employees can complete training.

To impact employees with diverse work roles, the team translated technical topics into plain language. S1 recognized the necessity of tailoring communications to employees' security knowledge and skill: *"I can sit up there and speak to the technical aspects of things and bore everybody except for the few techno-geeks in the back. Or I can try to make things more generalized" (S1)*. Employees expressed positive perceptions of the team's communication abilities. When asked about the content of the security awareness information, a staff member said it was beneficial *"having the information presented in a manner to where it's not intimidating,…in a way that you can embrace it and take away information" (E3)*.

## 5.2 Engaging and Empowering the Workforce

The agency's program transformation entailed a shift from an extrinsic motivation of security being an organizational compliance requirement to intrinsic motivation where security becomes a habit and something employees want to do because of its inherent value. S1 viewed effective security awareness as going beyond compliance: *"I'm not going to force anybody to change but give them the opportunity to see that they can."* S1 likened this goal to the development of dental hygiene habits:

> *"knowing that you have to brush your teeth, to actually brushing your teeth and then not even thinking about brushing your teeth every day. That's what I'm trying to push for in our program, where security awareness is now second nature" (S1).*

The shift involved changing workplace security culture by first facilitating recognition (engaging), then instilling a sense of personal responsibility and self-efficacy (empowering). The team recognized the importance of involving the workforce as active participants and contributors to the security of the organization, believing that *"Human beings are the most important cybersecurity tools" (S1)*. This section describes ways in which the team engaged and empowered the workforce.

### *5.2.1 Making Security Relatable*

Engagement was grounded in the personalization of the security message and bringing awareness of security issues and their relevance to individuals. At the observed security officer forum, S1 told the audience, *"We don't want to make you paranoid. We just want to make you aware" (S1)*. During the two observed security days, at the conclusion of each talk, S1 asked a question of each speaker to link the topic to Agency Q's mission and make the connection to attendees' jobs.

To make security relatable, the team included topical information during events, campaigns, and communications. Focus areas and event themes might be related to the season of the year or current news. For example, an event held shortly before the American football Super Bowl was entitled "Cyber Kick-off." Recent organizational security needs, threats to the agency, or hot security topics in the government also frequently drove topics. S1 provided an example:

> *"We just had an inspection where we found a bunch of people that were leaving their [smart cards] in their readers when they weren't at their desks. So, we're going to be talking at the forum about that" (S1).*

To ensure events were of interest, the team involved the workforce in deciding which topics to address via post-event surveys, personal interactions, email, and focus groups. The lead described how their team had increased attendance at the security officer forums:

> *"We brought a bunch of [security staff] together and asked, 'What do you want? What would make these more efficient and effective?'...We started bringing in people they wanted to hear, subjects they wanted to hear that meant most to them" (S1).*

Beyond topicality, the team strove to shift the mindset of employees to want to make informed security decisions and follow safe security practices regardless of context. This involved a shift of the locus of concern from just the security of the organization to also include the security of the individual. At most awareness events, there was a *"mix of cybersecurity at work and home" (S1)*. For example, at a holiday-themed lunchtime event, a table called "Information Station Holiday Edition: Are your gifts vulnerable?" was popular with attendees who learned about security and privacy considerations for devices like smart speakers and fitness trackers.

S2 commented that incorporating the home aspect gets people's attention because, at work, people might feel like someone else is responsible for security, but at home they are responsible. We asked employees how the information they received at work had helped them make security decisions at home. They discussed being able to recognize and act upon security issues regarding phishing emails, online shopping, and smartphone usage. An IT specialist appreciated the work-home emphasis:

> *"Some of the information, [the team] will pass out and say, 'You can send this and share this with family'...I definitely have thought about several of the talks going home. And I*

*have sent information that was approved to my parents saying, 'You need to read this'"* (E7).

The awareness events did not just focus on topics related to cybersecurity; rather, the team also tried to provide information related to other types of security affecting work and home life, such as personnel and physical security. The CISO commented on this approach: *"We want to secure the person, and we want everyone to think about all aspects in which they could secure themselves" (M1)*. For example, a summer "Traveling Cyber Safe" lunchtime event featured a personnel security officer providing information about foreign travel safety.

Of note, while the team endeavored to create engaging material and interactions, they acknowledged that their agency offered few tangible incentives for demonstrating good security behaviors (e.g., official recognitions or awards), in large part due to a lack of budget and staff available to progress any incentive initiatives. The agency was more apt to take negative measures, for example account suspension of employees who failed to complete their mandatory training.

### *5.2.2 Employing Engaging Communication Techniques*

To engage the workforce, the team disseminated security information using a variety of in-person and online communication techniques throughout the year to reinforce the message, rather than solely relying on once-a-year training. They took care in selecting topics of interest to the workforce and endeavored to find engaging, external speakers for security awareness events who were *"exciting for people to listen to. They're just hearing another perspective and another point of view that's not from our agency" (S2)*. To overcome complacency and garner attention while operating with a limited budget, the team recognized the need for creativity in their approaches:

> *"You want to just put a different spin on it because people just see stuff all the time: 'Have a good password. Lock your computer'…Be creative and think outside the box for different reasons or different tactics to make people think" (S3).*

A manager affirmed the need for security awareness efforts to creatively adapt to employees' preferences and constraints: *"We need to keep making it as user-friendly as possible, not having it be a big commitment of people's time. And doing it in such a way that people want to keep coming back to the program" (M1)*. As an example, during the agency's "Click with Care Campaign" that encouraged safe email behaviors, the team wanted to ensure that all members of the workforce viewed important anti-phishing tips. Hearing from employees that email reminders were typically not read, the team devised an alternative: *"We have a little phishing handout…We want to put it on half a piece of cardstock and go around at like 4:00 when everyone's gone and put it on everyone's desk with a little lifesaver [candy]" (S3)*.

The team focused on making security awareness memorable, often by holding events that were entertaining. Most lunchtime events included gamification, such as security-themed trivia with candy prizes. A team member discussed the team's frequent use of humor: *"If we can get five eye rolls at an event, we can call it a win. Because especially in this industry, everyone's so business and serious. So, we like to have a little fun" (S3)*. One creative approach was observed when the team executed a campaign to encourage employees to complete their annual security training early by distributing "Now and Later" candies on people's desks with postcards that said, "Take your training now and not later."

During the study timeframe, the team took the lunchtime events into new directions beyond the usual information fair setup. One observed event entitled "Late Night Cyberside Chat" was a humorous parody of a late-night television show. Various guests joined S1 "on stage" to discuss security and IT topics. Another observed event was Shakespeare-themed and entitled "To Send or Not to Send." The impetus for the theme stemmed from recent security issues observed within the agency in which employees were sending personally identifying information to their personal email accounts:

> *"We were brainstorming, and someone said, 'Oh, to send or not to send'…They're not paying attention to emails, they're not paying attention to posters. Maybe we'll do a little show about it and put our own cybersecurity spin on it" (S3).*

Attendees were given a playbill containing the performance script and security tips (see excerpts in Figure 1) and were offered a bag of popcorn. During the performance, S1 donned a Shakespearean-era hat and proceeded to read through three "acts" based off popular Shakespeare plays but re-written by the team to incorporate security themes.

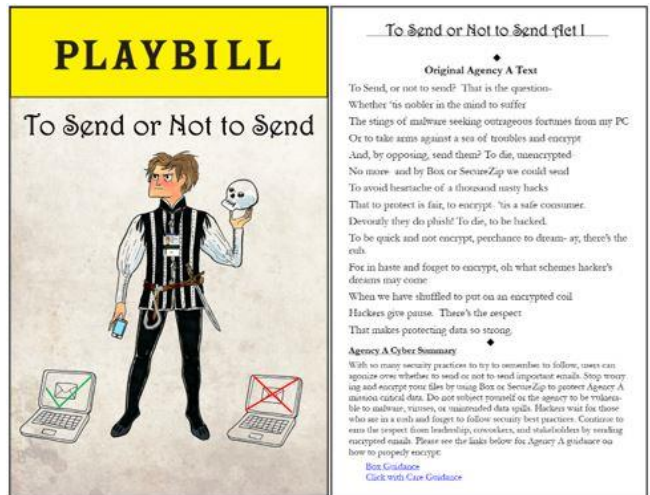

*Figure 1.* "To send or not to send" lunchtime event handout excerpts.
Original handout has been modified to anonymize the agency.

Although the team was not afraid to try new approaches, *"not everything works" (S2)*. The team members described a lunchtime event in which they designed a "cyber passport" for attendees:

> *"Everyone…could go and get a signature from each table [on the passport]…and put it in the box to win a gift card. And we thought it would be great, but it was just hard logistic-wise…And then a lot of people are kind of like 'Oh, I don't know I want to do this.' And when they found out it was a gift card to the cafeteria, they're like 'Oh, I don't want that.' …Not many people were filling them out and putting them in the box. So, that idea kind of fizzled" (S3).*

Less-successful events were viewed as learning experiences as they provided valuable information about workforce preferences.

### 5.2.3 Providing Practical Recommendations and Tools

Empowerment involves providing actionable steps, increasing feelings of self-efficacy, and encouraging individuals to be more reflective about their security behaviors. The team felt that bringing awareness of security threats was important but did not necessarily lead to behavior change. Therefore, people needed practical, actionable steps to counter threats and protect themselves and their organization. As expressed by S1, *"Cybersecurity awareness is not always 'The bad guys are coming to get you,' but 'Here are some better tools to use'."* S1 believed that people could take small steps that have a large, long-term impact: *"You need to just be aware of the little things you can do to protect yourself from the little things that are going to happen that are going to end up being a big pain" (S1)*. An employee commented on the synergy between bringing awareness and encouraging action:

> *"The more we know about it [security] and the more we know people that we can reach out to that can help us if we have a question about it, I think that can make you feel more empowered and more comfortable in doing it the right way and protecting yourself as well" (E3).*

The team tried to ensure awareness information included recommendations that were achievable given employees' skillsets, described in terms they understand, and were accompanied by points of contact for more information. Recommendations were often offered by event speakers and in security-related handouts provided at all events. For example, at the lunchtime "To Send or Not to Send" event, after each act, the team lead described the security issue in plain language and offered concrete steps attendees could take. The accompanying handout also included a list of agency resources for more information.

## 5.3 Measuring Success

Gauging the impact of security awareness training on workforce attitude and behavior change was essential for determining program effectiveness. The development of meaningful measures of effectiveness could be non-trivial but was still a goal for S1, who viewed compliance and effectiveness metrics as being complementary: *"I'm coming up with different ways of measuring how we're making that impact as well as making sure we're hitting all the right checkboxes for compliance. So, it's kind of a balancing act" (S1)*. Compliance served as a once-a-year indicator of coverage across the entire workforce while effectiveness measures provided continuous evaluation. Towards this ongoing assessment, the team utilized a combination of quantitative and qualitative measures.

### 5.3.1 Compliance Metrics

Compliance metrics revealed how many employees fulfilled their mandated annual security awareness training. These metrics were reported to agency leadership and government oversight organizations. Compliance, although sometimes deceiving, was one indicator of progress as expressed by the team lead: *"Seeing our training numbers, meeting our goals for the training numbers, even though that's just compliance, it's still showing that people realize that they have to take this, start making that awareness" (S1)*. Not meeting compliance goals could be disappointing. The team lead was upset when recent training compliance numbers were lower than desired due to an issue with employee rosters not being updated.

### 5.3.2 Event Attendance

Event attendance was an indicator of program popularity: *"I think we're starting to make a difference. I can see that by the numbers of people who come to our events and look forward to it" (S1)*. Although saying little about the impact, attendance was often the only immediately available measure. Since taking over the program, the team had seen an increase in attendance. Referring to security days, S1 compared the trend of 200-300+ attendees in the most recent three years with attendance prior to the team becoming involved: *"We've also seen some after action reports from some of the other cybersecurity events that took place…They were excited to get 50 people."* In addition to event attendance, there was an increase in the number of people who viewed online posted content after the event. Whether because of improvements in quality or better marketing of events, the team and their leadership interpreted these numbers as a positive sign.

Event attendance was not just assessed by number of attendees but also by who attended as an indicator of reach across the agency. The team had observed an increasingly diverse audience beyond those with security roles. However, there was still a large portion of the agency population who chose not to attend events.

To increase reach, the team was considering new approaches. Because employees may be reluctant to step away from work to attend events, ideas for future improvement centered on the team visiting staff. M2 suggested that the team consider occasionally attending group meetings to talk about security. Because of thin resources, M2 also discussed the possibility of having security "ambassadors" within each mission organization. These ambassadors could serve as an extension of the security awareness team by passing on security tips and reminders to coworkers.

### *5.3.3 Employee Feedback*

Employee feedback was another way in which the team gauged the effectiveness of their program. After every security day and security officer forum, the team invited attendees to complete an anonymous, post-event survey. Although results may be subject to self-selection bias, surveys provided a useful mechanism to obtain feedback. Respondents rated the quality of topics and presentations, event organization, and communications and could provide additional feedback and suggestions for future events. Overall, survey respondents viewed the events positively. Out of 43 respondents in two security officer forum surveys, 93% rated the quality of speakers and topics as above average or excellent. Out of 109 respondents in three security day event surveys, 95% rated the quality as above average or excellent. For the second security day we observed, the survey additionally asked why people attended the event. Out of the 21 respondents, two-thirds indicated that they wanted to earn training credit, while over 85% said that the advertised talks looked interesting. Just over half attended because they thought the information would be helpful in their jobs. These reasons provided insight into employee motivations for attending and their perceived value of the events.

In addition to survey responses, the team often received feedback directly from employees (face-to-face, email). Although not quantifiably measurable, direct feedback provided anecdotal evidence that security awareness was shifting security attitudes and generating interest. Staff members told S3 how much they enjoyed past events: *"We were still hearing things about speakers we had a couple years ago, and people saying, 'Oh can we get them back, can we get someone similar?'…The fact that we still hear feedback months and years later is very rewarding" (S3)*. Even managers in the team's chain of command received feedback: *"I'm getting a lot more positive feedback from the agency, from the user community, than I had in the past. And before, when we thought not getting negative feedback was good, getting positive feedback is even better" (M1)*.

Our interviews with agency employees also shed light on the program's impact. Several employees with security roles commented on the value of information presented at the security officer forums, with an IT specialist mentioning that one talk *"made me think more about my system... [It] was something that I could apply to my own systems going forward" (E7)*. Over half the participants discussed the personal impact of awareness efforts related to phishing and appropriate email behaviors. An employee commented, *"I didn't have that awareness before… I am much more educated now than I used to be… That by itself is a big deal" (E4)*. Others demonstrated that their awareness also translated into action, for example, being empowered to report potential phishing emails. An attorney remarked on how agency security awareness efforts led to increased vigilance:

> *"You can get messages from the agency, and I go like, 'Is this real?' And I send it to the cybersecurity team to say, 'Am I supposed to be answering this?' So, basically, as a result of my training here, I am very, very suspicious of everything" (E5).*

### *5.3.4 User-Generated Incidents and Reporting*

Trends in employee-involved security incidents served as additional evidence that the workforce was becoming more security-aware: *"Are they [incidents] going up? Are they coming down? Are we seeing more people going to the training and does that mean we're seeing fewer events?" (S1)*. For example, indicators of behavior change might include decreases in the number of incidents of badges being left unattended in computers or fewer incidents of employees emailing unencrypted personally identifiable information.

A decrease in phishing exercise click rates (number of employees clicking on the phishing email link) and number of repeat clickers (those who click on phishes in two or more consecutive exercises) also provided useful information. When click rates dipped lower, S1 wondered if this was because the phishes were too easy or if the security culture was improving. To test this, the team increased the sophistication of phishes sent during quarterly exercises. The subsequent click rates remained low, implying that employees were indeed informed about phishing and making sound decisions.

Employee reporting of security issues was also viewed as a success indicator. When asked what measures of effectiveness were most meaningful, the CISO commented, *"Once upon a time, I would have said fewer incidents. Now, I'm saying more better-reported incidents. So, people are recognizing that they've done something, recognizing that there's a problem" (M1)*. For example, the agency saw an uptick in phishing reporting for both phishing exercises and suspected real-word phishing emails.

S1 wanted to take a more holistic approach to measuring program effectiveness. S1 expressed the desire to work with security staff to examine agency security data (e.g., logs, incident reporting, real-world phishing emails) and physical security data (e.g., badge incidents) to explore potential relationships to awareness efforts and identify areas where new or different training might be beneficial. Phishing was discussed as an exemplar of how a holistic approach might work:

> *"We're trying to…be able to tie in together the people who take their training to the people who get caught with phishing exercises, the people who are really getting caught with phishing exercises with people who are losing their badges to people who send out information they shouldn't to see what's the correlation here. Are these people just too busy? Are they not paying attention? Is there a training problem?" (S1)*.

Unfortunately, despite having solid relationships with other agency security groups, the team had not been able to make much progress on these goals because of limited resources.

### *5.3.5 Evidence of Leadership Support*

Another indicator of success was the increase of support from senior leaders. S1 believed, *"If we can get the leadership to look forward to what we're doing and show some interest rather than just being another line on a report, I think that's very good. I think we're making some progress" (S1)*. In years past, the leadership viewed the program as *"just kind of this small little program. I don't think it had that much spotlight on it" (S3)*. However, that had changed: *"Now I've got the CIO and the deputy CIO and the CISO and the deputy CISO sitting in the crowd and watching the whole [event] rather than just showing up to give their opening remarks and then leaving" (S1)*. The interviewed managers were pleased with the progress the team had made. S1's immediate supervisor complimented, "*I think we're doing a fantastic job with the resources we have" (M2)*. S1 also engaged with senior leadership outside their chain of command: *"I also do the senior executive training, and I'm trying to make them more aware of what we're doing and why we're doing it. And so far, I've gotten a lot of positive feedback" (S1)*.

# 6 Discussion

Prior security awareness research studies predominantly explored impacts on the recipients of training or via self-report data. However, our case study goes beyond these to holistically uncover work practices of security awareness professionals by observing awareness activities in action and exploring program transformation over a prolonged time. We further supplement our observations by uniquely synthesizing multiple stakeholder perspectives. In this section, we situate our study in the context of prior literature and describe its real-world implications for security awareness programs.

## 6.1  Progressing Security Awareness Programs

The case study further contributes to the research body of knowledge as well as providing practical implications for organizational security awareness programs.

### 6.1.1  Advancing Security Awareness Research

Our findings provide observable evidence that validates the challenges of security awareness programs identified in prior self-report surveys and studies. For example, Agency Q's security awareness team sometimes struggled to overcome negative perceptions of awareness training (Bada *et al.*, 2019; Alshaikh, 2018) and faced resource issues exhibited by a constrained budget and S1 performing awareness duties part-time (SANS, 2021; Woelk, 2015).

The team attempted to overcome these challenges in a real-world demonstration of the approaches previously recommended (but not often observed) by researchers and industry experts. Approaches included: reinforcing security awareness throughout the year (Bada et al., 2019); tailoring communications to the audience's knowledge and emotional response to security; soliciting and acting upon employee feedback (Alshaikh *et al.* 2018); developing creative and varied ways to engage the workforce (Bauer *et al.*, 2017; Abawajy, 2014; Korpela, 2015); collecting a variety of measures of effectiveness (Alshaikh *et al.*, 2018; Fertig *et al.*, 2020); and gamification (Khando et al., 2021; Hart et al., 2020).

Our case study also confirms and provides real-world examples of the habit, fear, and role value elements identified in Moody et al.'s UMISPC (2018). The team facilitated *habit* by providing information applicable to both work and home so that good security behaviors become ingrained in daily life. *Fear* was approached by presenting honest perspectives on security threats while tempering that by building employee self-efficacy via concrete guidance and resources. The team addressed *role values* by tailoring and making their communications relatable to different employee groups. We further discovered real-world implementation of the three dimensions of security awareness identified by Hänsch and Benenson (2014). The team facilitated *perception* by providing information about specific security threats and *protection* by providing practical recommendations. *Behaviors* were reflected to a certain degree by phishing simulation metrics and employee feedback; however, the team acknowledged that they have more work to do to gather and synthesize additional behavior-based measures.

Agency Q's team did little in the way of providing external, positive incentives, such as employee recognitions or competitions, which have been anecdotally cited as helpful in promoting positive security behaviors and attitudes (Haney et al., 2022). However, other researchers found that such rewards had no significant impact on security behavior intentions (Moody et al., 2018). Regarding punishments, the agency suspended accounts of employees who failed to complete annual training. While effective in the short term to achieve compliance, prior work has shown that punishments and excessive fear appeals can be counterproductive in establishing long-term positive attitudes towards security, so should be used carefully (Renaud and Dupuis, 2019).

*6.1.2   Practical Implications for Programs*

Our findings can serve as a resource to guide the work of security awareness professionals in transforming their own programs. While acknowledging that these approaches may not be suitable for all organizations, our observations afford a unique opportunity to witness how a security awareness team goes about rethinking and progressing their program.

As a basis of comparison for other organizations, we situate this transformation along the five increasing levels of maturity in the SANS Security Awareness Maturity Model (2018):

- Level 1, Non-existent program
- Level 2, Compliance-focused - program focused on meeting annual training requirements
- Level 3, Promoting Awareness and Behavior Change - program includes topics that directly impact the organization's mission; awareness activities conducted continuously throughout the year
- Level 4, Long-term Sustainment and Culture Change - program has processes, adequate resources, leadership support, and positive impact on security culture
- Level 5, Metrics Framework - program collects metrics, including behavioral impact, resulting in continuous improvement and demonstration of return on investment

The case study revealed specific examples of how a security awareness program can evolve beyond Level 2 (found in many organizations) by translating the intent of the training policy into organization-relevant approaches aimed at changing attitudes and behaviors of individuals. For example, the team demonstrated meeting the threshold for Level 3 by picking topics relevant to the organization and its employees. The team also distributed security information throughout the year using a variety of engaging formats – informal drop-in events, structured security days, and physical materials – to accommodate differences in employee preferences. The program's use of gamification and humor further attracted attendees to events, embodying security expert Adam Shostack's (2023) view that "Security is very serious stuff, but that doesn't mean that we can't be playful, creative, or engaging as we work."

Through its efforts, the program further demonstrated most elements needed to advance to Level 4 by facilitating shifts in organizational security culture, garnering more leadership support, and starting to gauge behavior change by collecting a variety of measures. However, they still struggled in that some employees did not see how security relates to their jobs.

The agency program had not yet attained the highest maturity level, as the team recognized that there were still challenges and improvement opportunities, largely due to resource shortfalls. They implemented some of recommended metrics in Chaudhary *et al.*'s framework (2022). The team collected accessibility indicators to measure relevance, quality, and reachability via employee feedback surveys and attendance reports of which groups most frequently attended events.  Monitoring indicators to gauge organizational support came via surveys, leadership involvement, and informal feedback. They implemented impact indicators in the form of phishing click rates; however, security incidents were not correlated to the extent they would have liked. The team also did not measure sustainability indicators of the impact on organizational policies and program funding over time.

We also note another shortfall that can serve as a lesson learned for other programs: the lack of utilization of established instructional design principles or learning (e.g., Experiential Learning Theory) (Aaltola and Taitto 2019) and behavioral theories (e.g., Protection Motivation Theory, Theory of Planned Behavior) (Moody et al., 2018; Lebek et al., 2014), which can be valuable in informing awareness approaches. Part of this issue may be attributed to the research-practice gap common to many disciplines. In addition, the lack of involvement of learning experts and communications and marketing staff – whose practices are based on enduring behavioral theories – may put the agency team at a disadvantage.

*6.2 Progressing the Security Awareness Role*

Our findings provide additional insights and understanding of security awareness professionals as a unique work role.

*6.2.1 Advancing Research on the Security Awareness Role*

Program transformation was facilitated by the security awareness team transitioning from being compliance managers to becoming the change agents, risk communicators, and cybersecurity advocates identified in prior work. We extend prior work that outlined desired knowledge and skills for these professionals by illustrating – through direct observations and multi-stakeholder perspectives – these skills and how they are perceived by others in a real-world context.

The team displayed not just technical acumen but also non-technical, professional skills (e.g., interpersonal skills, listening, communication skills, context awareness) previously identified as necessary to establish trust and credibility with recipients of their communications (Covello, 1997; Slovic, 1987; Rogers, 2003; Nurse et al., 2011; Haney and Lutters, 2021; Haney et al., 2022). For example, the team regularly solicited employee feedback, were viewed as friendly and approachable, and took special pains to communicate in plain language. They also had the multi-disciplinary composition common in advocacy (Haney and Lutters, 2021) and security awareness (Haney et al., 2022) teams. Although only the team lead had a technical background, the skills brought to bear by the other two team members (e.g., interpersonal, creative, planning) contributed to the program's progression and demonstrated the benefit of having diverse skillsets in awareness teams.

*6.2.2 Practical Implications for Supporting the Security Awareness Role*

Evidence of professional competencies can contribute to industry efforts to formally define a security awareness role within security work role frameworks, including the widely-adopted National Initiative for Cybersecurity Education Workforce Framework for Cybersecurity (NICE Framework) (Petersen *et al.*, 2020). The NICE Framework includes Work Roles consisting of tasks, knowledge, and skills. Private and public sector organizations have utilized these Work Roles to hire security workers, build teams to achieve specific objectives, shape career paths, and discover critical workforce gaps (NIST, 2020).

Currently, the Framework lacks a Work Role that adequately captures the duties and necessary knowledge and skills of security awareness professionals (Haney et al., 2022). To remedy this, SANS (2019) proposed a new Work Role called "Security Awareness and Communications Manager," which places less emphasis on technical skills and more on professional skills such as communications, partnering, and project management. Spurred in part by the SANS proposal, our own research, and other community feedback, the NICE Community Council hosted a workshop to discuss the need for a new awareness-focused role (NICE, 2021).

# 7 Conclusion

This paper describes a longitudinal case study of how one government agency's security awareness program has transformed from being merely compliance-focused to a higher degree of maturity involving sustained empowerment and engagement of organizational employees. The discoveries described in this paper contribute to the research body of knowledge on security workers who aim to facilitate security behavior change by providing evidence from a real-world context, validated by the perspectives of multiple stakeholders. In addition, the tactics, professional characteristics, and program progression demonstrated by the agency team can serve as a valuable exemplar and resource for security awareness professionals in other organizations as well as industry and government groups seeking to better define security work roles.

## Disclaimer

Certain commercial companies or products are identified in this paper to foster understanding. Such identification does not imply recommendation or endorsement by the National Institute of Standards and Technology, nor does it imply that the companies or products identified are necessarily the best available for the purpose.

## Acknowledgements

We would like to thank the security awareness team, office of the Chief Information Officer, and the employees at Agency Q who were so generous with their knowledge and time and afforded us this research opportunity.